\documentclass[superscriptaddress,reprint,aps,pra]{revtex4-1}
\usepackage{graphicx}
\usepackage[margin=0.7in]{geometry}
\usepackage{amsmath}
\usepackage{hyperref}
\usepackage{multirow}
\usepackage[usenames]{color}

\let\hatOrig\hat
\renewcommand{\vec}[1]{\boldsymbol{\mathbf{#1}}}
\renewcommand{\hat}[1]{\boldsymbol{\mathbf{\hatOrig{#1}}}}
\renewcommand{\Im}{\operatorname{Im}}
\renewcommand{\Re}{\operatorname{Re}}
\newcommand{\D}{\mathrm{d}}
\newcommand{\sub}[1]{\ensuremath{_{\textrm{#1}}}}   
\begin{document}

\title{Hot carrier dynamics in plasmonic transition metal nitrides}

\def\RPIPhy{Department of Physics, Applied Physics and Astronomy, Rensselaer Polytechnic Institute, Troy, NY 12180, USA}
\def\RPIMSE{Department of Materials Science and Engineering, Rensselaer Polytechnic Institute, Troy, NY 12180, USA}
\def\EqualContrib{These authors contributed equally}

\author{Adela Habib}\thanks{\EqualContrib}\affiliation{\RPIPhy}
\author{Fred Florio}\thanks{\EqualContrib}\affiliation{\RPIPhy}
\author{Ravishankar Sundararaman}\email{sundar@rpi.edu}\affiliation{\RPIPhy}\affiliation{\RPIMSE}
\date{\today}

\begin{abstract}
Extraction of non-equilibrium hot carriers generated by plasmon decay in metallic
nanostructures is an increasingly exciting prospect for utilizing plasmonic losses,
but the search for optimum plasmonic materials with long-lived carriers is ongoing.
Transition metal nitrides are an exciting class of new plasmonic materials with
superior thermal and mechanical properties compared to conventional noble metals,
but their suitability for plasmonic hot carrier applications remains unknown.
Here, we present fully first-principles calculations of the plasmonic response,
hot carrier generation and subsequent thermalization of all group IV, V and VI
transition metal nitrides, fully accounting for direct and phonon-assisted
transitions as well as electron-electron and electron-phonon scattering.
We find the largest frequency ranges for plasmonic response in ZrN, HfN and WN,
between those of gold and silver, while we predict strongest absorption
in the visible spectrum for the VN, NbN and TaN.
Hot carrier generation is dominated by direct transitions for most of the
relevant energy range in all these nitrides, while phonon-assisted processes
dominate only below 1 eV plasmon energies primarily for the group IV nitrides.
Finally, we predict the maximum hot carrier lifetimes to be around 10 fs
for group IV and VI nitrides, a factor of 3 - 4 smaller than noble metals,
due to strong electron-phonon scattering.
However, we find longer carrier lifetimes for group V nitrides, comparable
to silver for NbN and TaN, while exceeding 100 fs (twice that of silver) for VN,
making them promising candidates for efficient hot carrier extraction.
\end{abstract}

\maketitle

\section{Introduction}

Transition metal nitrides (TMNs) have a long history of applications in cutting tools
and corrosion resistance coatings \cite{HardCutting,corrosionResistantCoating}
based on their desirable mechanical properties, especially their abrasion
resistance and high melting points \cite{Toth, TiNAlloyAbrasiveResistent}.
They have also attracted interest for their electronic properties, with applications
including diffusion barriers \cite{TMNdiffusionbarrier,TiNZrNDiffusionBarrier}, 
metallization materials \cite{MetalizationMaterial},
field emitter cathodes \cite{NbNFieldEmitterCathod} and
even as high temperature superconductors \cite{Superconducting}.
Most recently, TMNs are receiving considerable attention as general-purpose
and `refractory' plasmonic materials for optical-frequency applications
at ambient and elevated temperatures on account of their high melting points,
high electron mobilities and lower cost than noble metals \cite{refractoryPlasmonics,TiNAsPlasmonicMatExp,PlasmonicNitridesReview,e-ph-IV}.

Plasmonic resonances of metal surfaces and nano-structures hold the exciting promise
of shrinking optical and photonic devices from the micro- to the nano-scale,
with applications including photodetection, communication, computing and biosensing
\cite{chipScalePlasmonicApplication,biosensingPlasmonApp}.
Where stability of nanostructured plasmonic materials in harsher environments
(chemical) becomes a necessity, TMNs are increasingly replacing
noble metals \cite{BioMedicalTiNPlasmonApp,TiNPlasmonicTechExplain}.
A central challenge in plasmonic applications is that plasmons
rapidly decay to electron-hole pairs in the material, but that loss
mechanism can be gainfully exploited as a pathway for converting photons
to energetic `hot' carriers that can be extracted for solar energy conversion,
photochemistry and photodetection \cite{hotCarrierApplications1,hotCarrierPhotoChemistry}.
The challenge then becomes extracting hot carriers before they thermalize by scattering
against phonons and other carriers in tens of femtoseconds \cite{hotCarrierTheoryReview}.
Efficient hot carrier devices require new materials with long lifetimes and transport
distances for high-energy carriers, a subject of active current research \cite{GraphiteHotCarriers}.

Transition metal nitrides could potentially surpass noble metals such as gold for broad-band
photothermal as well as narrow-band applications in the visible region, with TiN predicted to be
the most efficient solar spectrum absorber among them \cite{PlasmonicResponseTheoryTMN}.
While considerable recent attention to these materials has focused on their optical
and plasmonic response, less is known so far about details of non-radiative plasmon decay,
hot carrier generation mechanisms and their subsequent relaxation dynamics in these materials.
Detailed theoretical and experimental studies of hot carrier dynamics in TMNs,
analogous to those in noble metals \cite{DirectTransitions,PhononAssisted,MultiPlasmon,
TAparameters,TAanalysis}, are now necessary to determine the real potential of these
exciting materials for plasmonic hot carrier applications.

In this article, we present \emph{ab initio} calculations of the optical response,
hot carrier generation from direct and phonon-assisted transitions, and carrier
relaxation by electron-electron and electron-phonon scattering of nine
plasmonic TMNs: the nitrides of all group IV, V and VI transition metals.
We calculate all optical, electron, and phonon interactions fully from 
first-principles, with no experimentally-determined parameters, allowing
us to predict the potential of the ideal materials free of defects.
In particular, we predict the theoretical resistivities of all these materials,
which we find below to be in excellent agreement with measurements on single-crystalline
stoichiometric samples, currently available only for group IV nitrides.
By analyzing the generated carrier distributions and subsequent thermalization lifetimes,
we show below that the group V nitrides are particularly promising for plasmonic
hot carrier applications. Specifically, we find carrier lifetimes in VN can
exceed those in the lowest-loss noble metal, silver, by more than a factor of two.

\section{Methods}

We calculate equilibrium structures, electron and phonon band structures,
and electron-optical and electron-phonon matrix elements for the nine
group IV, V and VI nitrides using the open-source plane-wave density
functional theory (DFT) software, JDFTX \cite{JDFTx}.
We use the `PBE' \cite{PBE} generalized-gradient approximation to the exchange-correlation functional,
norm-conserving `SG15' \cite{SG15} pseudopotentials, a kinetic energy cutoff of 30 $E_{h}$ (Hartrees)
and a Fermi-Dirac smearing of $0.01~E_h$.
For Brillouin zone integration, we employ a $\Gamma$-centered $k$-point mesh 
with dimensions such that the effective supercell is nominally 25~\AA~ in each direction
for the electrons, while for the phonons, we use a supercell nominally 12~\AA~ in each direction.
(This amounts to a $8\times 8\times 8$ electronic $k$-mesh and a $4\times 4\times 4$
phonon $q$-mesh for the simplest rock-salt and hexagonal structures below,
which then reduce proportionally for the larger unit cells.)

We construct maximally-localized Wannier functions \cite{MLWFmetal} that exactly
reproduce the DFT band structure up to at least 5~eV from the Fermi level,
starting from a trial set of metal $d$ orbitals, N $p$ orbitals and
Gaussian orbitals at the largest void sites of each crystal structure.
We convert all electron and phonon properties to a localized basis
using this representation, and use that to efficiently evaluate them at
arbitrary electron and phonon wave-vectors ($k$ and $q$ respectively)
for evaluating excitation and scattering rates using Fermi's Golden rule,
as we discuss in the Results section for each property separately.
In particular, we use this basis to directly evaluate
electron energy eigenvalues $\varepsilon_{\vec{k}n}$ with band index $n$
and corresponding Fermi occupation factors $f_{\vec{k}n}$,
phonon frequencies $\omega_{\vec{q}\alpha}$ with mode index $\alpha$
and corresponding Bose occupation factors $n_{\vec{q}\alpha}$,
electron velocity matrix elements $\vec{v}^{\vec{k}}_{n'n}$,
and electron-phonon matrix elements $g^{\vec{q}\alpha}_{\vec{k}'n',\vec{k}n}$
(with $\vec{q}=\vec{k}'-\vec{k}$ for crystal momentum conservation).
This approach allows us to directly evaluate all physical properties
using \emph{ab initio} matrix elements sampled over the entire Brillouin zone,
rather than a simplified treatment that relies on deformation potentials
or that includes only certain phonon branches or special phonon modes.
See Refs.~\citenum{PhononAssisted} and \citenum{GraphiteHotCarriers}
for more details.

\section{Results and Discussion}

\subsection{Structure}

\begin{figure}
{\centering\includegraphics[width=0.8\columnwidth]{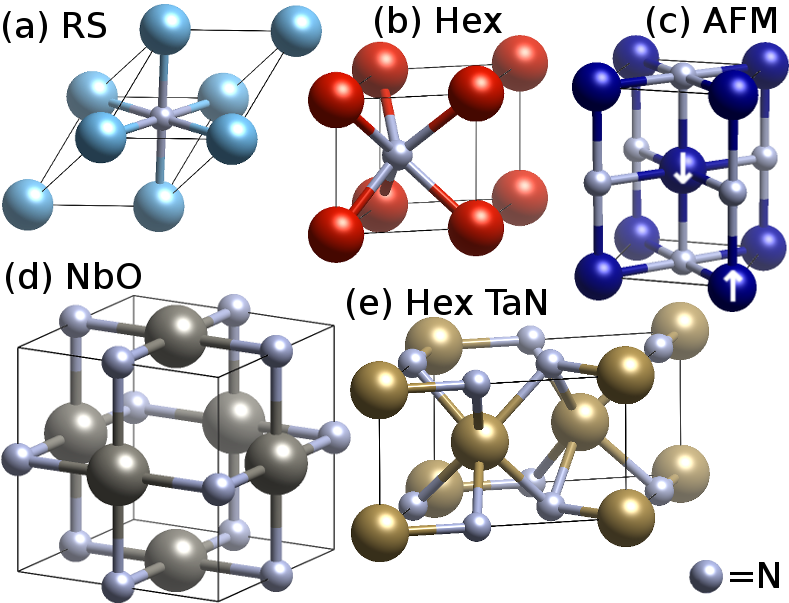}}
\caption{Primitive unit cells for: (a) rock-salt structure of TiN, ZrN, and HfN,
(b) hexagonal structure of VN, NbN, CrN, MoN, and WN,
(c) anti-ferromagnetic structure for CrN,
(d) NbO structure for MoN and WN, and
(e) hexagonal structure for TaN.
\label{fig:Structures}}
\end{figure}

\begin{table}
\caption{Structure and stability of transition metal nitride phases.
Stable structures are rock salt (RS) for group 4 nitrides,
hexagonal (Hex) for group 5, and cubic NbO for group 6,
except CrN which is antiferromagnetic with a tetragonal distortion of the RS structure.
Thermodynamic stability is indicated by energy differences,
while mechanical stability by finite zero-point energy (ZPE);
ZPE = N/A indicates imaginary phonon frequencies and mechanical instability.
\label{tab:structure}}
\begin{tabular}{|c|c|cc|ccc|c|c|}
\hline
\multirow{2}{*}{TMN} & \multirow{2}{*}{Phase}
	& \multicolumn{2}{c|}{Calc [\AA]} & \multicolumn{3}{c|}{Expt [\AA]}
	& \multirow{2}{*}{$\Delta E$ [eV]} & \multirow{2}{*}{ZPE [eV]}\\
& & $a$ & $c$ & $a$ & $c$ & Ref & & \\
\hline
TiN & RS & 4.253 & & 4.242  & &\cite{TiNLatticeExp} & 0.00 & 0.139 \\
\hline
ZrN & RS & 4.599 & & 4.578 & &\cite{ZrNLatticeExp}  & 0.00 & 0.117 \\
\hline
HfN & RS & 4.551 & & 4.524 & & \cite{HfNLatticeExp} & 0.00 & 0.112 \\
\hline
\hline
\multirow{2}{*}{VN}
 & RS  & 4.129 &       & 4.132 & & \cite{VNcubicLat} & 0.24 & N/A \\
 & Hex & 2.747 & 2.644 & & & & 0.00 & 0.172 \\
 & NbO & 3.952 & & & & & 1.11 & N/A \\
\hline\multirow{2}{*}{NbN}
 & RS  & 4.432 & & 4.392 & & \cite{NbNcubicLat} & 0.28 & N/A \\
 & Hex & 2.966 & 2.881 & & & & 0.00 & 0.140 \\
 & NbO & 4.263 & & & & & 0.93 & N/A \\
\hline
\multirow{2}{*}{TaN}
 & RS  & 4.430 & & 4.338 & & \cite{TaNcubicHexLat} & 0.61 & N/A \\
 & Hex & 5.225 & 2.924 & 5.185 & 2.908 & \cite{TaNcubicHexLat} & 0.00 & 0.141 \\
 & NbO & 4.253 & & & & & 0.81 & N/A \\
\hline
\hline\multirow{2}{*}{CrN}
 & AFM & 4.147 & 2.938 & & & & 0.00 & 0.122\\
 & Hex & 2.698 & 2.599 & & & & 0.00 & 0.156 \\
 & NbO & 3.833 & & & & & 0.55 & 0.135 \\
\hline\multirow{2}{*}{MoN}
 & RS  & 4.347 & & & & & 0.59 & N/A \\
 & Hex & 2.877 & 2.852 & & & & 0.06 & 0.113 \\
 & NbO & 4.112 & & & & & 0.00 & 0.127 \\
\hline\multirow{2}{*}{WN}
 & RS  & 4.368 & & & & & 1.38 & N/A \\
 & Hex & 2.870 & 2.906 & & & & 0.61 & 0.104 \\
 & NbO & 4.119 & & & & & 0.00 & 0.128 \\
\hline
\end{tabular}
\end{table}

Stoichiometric nitrides with formula MN, where M is a transition metal from
groups IV-VI are most often experimentally synthesized and found to be stable
at room temperature in the rock-salt (RS) structure (Fig.~\ref{fig:Structures}(a))
\cite{TiNLatticeExp,HfNLatticeExp,ZrNLatticeExp,VNcubicLat,NbNcubicLat,TaNcubicHexLat}.
Table~\ref{tab:structure} summarizes our DFT calculations of energies and
mechanical stabilities of RS and other structures for these nitrides.
For group IV nitrides, the RS structure is predicted to be stable and
the calculated lattice parameters are in good agreement with experiment
\cite{TiNLatticeExp,ZrNLatticeExp,HfNLatticeExp} (within 1~\%).
In this case, the mechanical stability of the structure is indicated by
the calculated phonon spectra (shown in SI) containing no imaginary frequencies,
which is summarized in Table~\ref{tab:structure} by a positive well-defined zero-point energy (ZPE).

However, we find the experimentally prevalent RS structure to be mechanically unstable
for nitrides of groups V and VI, indicated by imaginary frequencies in the phonon dispersion
relations (in the SI) and an undefined ZPE in Table~\ref{tab:structure}, in agreement with
previous theoretical calculations on these materials \cite{NitridesAndCarbidesImaginaryFreqPhonon}.
The primary reason for this discrepancy is that DFT calculations predict
energies and forces at zero temperature, and it turns out for these materials
that the room-temperature stable phases are unstable at zero temperature.
Specifically for VN, the RS structure is unstable towards a tetragonal distortion below 250~K,
and is only stabilized above this temperature by anharmonic lattice vibrations.\cite{VNcubicLat}
Predicting electron-phonon and optical properties, fully from first principles,
for phases that are mechanically unstable at zero temperature is not currently
practical with available methods, and will be a subject of future work.
Instead, we will focus here on zero-temperature-stable phases of these materials.

We find that all group V and VI nitrides have a mechanically stable hexagonal (Hex)
phase (Fig.~\ref{fig:Structures}(b)) with a lower energy than the unstable RS structure.
Specifically, for TaN, the stable structure is the more complex hexagonal structure
shown in Fig.~\ref{fig:Structures}(e), in good agreement with experiment
\cite{TaNcubicHexLat} with predicted lattice parameters within 1~\%.

The group VI nitrides have an additional complication, where the experimentally observed
RS structures, for example in WN, could be stabilized by vacancy formation \cite{WNstabilityGall}.
Once again, performing \emph{ab initio} electron-phonon and optical response calculations
on large unit cells with a low concentration of vacancies is currently impractical.
Instead, following Ref.~\citenum{WNstabilityGall} we focus on a symmetric
high-concentration limit (25~\%) of ion-pair vacancies in the RS structure,
which results in the NbO structure shown in Fig.~\ref{fig:Structures}(d).
We find this structure to be mechanically stable for all group VI nitrides,
and to be the lowest energy structure for MoN and WN.
It is however not stable for group V nitrides, for which only the Hex phases are found to be stable at zero temperature.

Finally, for CrN, the RS structure is stabilized by a tetragonal distortion
with anti-ferromagnetic (AFM) order, as shown in Fig.~\ref{fig:Structures}(c)
and is in agreement with previous theoretical predictions \cite{CrNImaginaryFreqPhonon}.
This structure is very close in energy to the Hex structure,
and lower in energy than the NbO structure.
We also find that all structures other than the RS CrN prefer to remain non-magnetic.
We perform \emph{ab initio} plasmonic response and hot carrier calculations for all
the mechanically stable structures discussed above: the RS structure for group IV,
the Hex structures for group V, and the Hex and NbO structures for group VI nitrides,
except CrN, where we focus on the two low-energy Hex and AFM structures.

\subsection{Resistivity}

We start with \emph{ab initio} calculations of the resistivity;
the near-equilibrium transport of charge carriers in these materials,
which is also an important constituent of the optical / plasmonic response below.
We use a linearized Boltzmann equation with a full-band relaxation-time approximation
\cite{PhononAssisted,MobilityRTE}, where the conductivity tensor is calculated as
\begin{equation}
\bar{\sigma} \equiv \bar{\rho}^{-1} = \int\sub{BZ}\frac{e^2\D\vec{k}}{(2\pi)^3} \sum_n
	\frac{\partial f_{\vec{k}n}}{\partial \varepsilon_{\vec{k}n}}
    (\vec{v}_{\vec{k}n}\otimes\vec{v}_{\vec{k}n})
    \tau^p_{\vec{k}n},
\label{eqn:sigma}
\end{equation}
where the integral over the Brillouin Zone (BZ) is carried out using Monte Carlo
sampling with $\sim 10^7$ $k$ values, with all terms in the integrand calculated
efficiently using the Wannier representation as discussed in the Methods section.
Above, the Fermi occupation energy derivative selects carriers near the Fermi level,
the band velocities $\vec{v}_{\vec{k}n}$ are just the diagonal terms in
the aforementioned velocity matrix elements ($\vec{v}^{\vec{k}}_{nn}$)
and $\otimes$ denotes tensor / outer product.
The final term $\tau^p_{\vec{k}n}$ is the momentum relaxation time,
whose inverse (momentum scattering rate) is given by Fermi's golden rule as
\begin{multline}
(\tau^p_{\vec{k}n})^{-1} =
\frac{2\pi}{\hbar} \int\sub{BZ} \frac{\Omega d\vec{k}'}{(2\pi)^3} \sum_{n'\alpha\pm}
	\delta(\varepsilon_{\vec{k}'n'} - \varepsilon_{\vec{k}n} \mp \hbar\omega_{\vec{k}'-\vec{k},\alpha})
\\
\times
	\left( n_{\vec{k}'-\vec{k},\alpha} + \frac{1}{2} \mp \left(\frac{1}{2} - f_{\vec{k}'n'}\right)\right)
	\left| g^{\vec{k}'-\vec{k},\alpha}_{\vec{k}'n',\vec{k}n} \right|^2 \\
\times
	\left(1 - \frac{\vec{v}_{\vec{k}n}\cdot\vec{v}_{\vec{k}'n'}}{|\vec{v}_{\vec{k}n}| |\vec{v}_{\vec{k}'n'}|}\right),
\label{eqn:tauInv_ePhP}
\end{multline}
where the terms in the product, in order, represent energy conservation, occupation factors,
the electron-phonon matrix elements and the angle between initial and final electron velocities
after scattering. The sum over $\pm$ accounts for phonon absorption and emission processes,
and we point out once again that all quantities above originate from DFT via the Wannier representation.
See Ref.~\citenum{PhononAssisted} and \citenum{GraphiteHotCarriers} for further details.

To ease analysis, we also define two averaged quantities not directly used in the resistivity calculation above.
First, we define the average Fermi velocity by
\begin{equation}
v_F^2 = \frac{
\int\sub{BZ}\frac{\D\vec{k}}{(2\pi)^3} \sum_n 
	\frac{\partial f_{\vec{k}n}}{\partial\varepsilon_{\vec{k}n}} |\vec{v}_{\vec{k}n}|^2
 }{
\int\sub{BZ}\frac{\D\vec{k}}{(2\pi)^3} \sum_n 
	\frac{\partial f_{\vec{k}n}}{\partial\varepsilon_{\vec{k}n}}
},
\label{eqn:vF}
\end{equation}
where the Fermi occupation derivative selects all states near the Fermi level
that contribute to conduction (within a few $k_BT$ of $\varepsilon_F$).
Second, we define the average `Drude' relaxation time,
\begin{equation}
\tau_D =\frac{
\int\sub{BZ}\frac{\D\vec{k}}{(2\pi)^3} \sum_n 
	\frac{\partial f_{\vec{k}n}}{\partial\varepsilon_{\vec{k}n}} |\vec{v}_{\vec{k}n}|^2 \tau_{\vec{k}n}
 }{
\int\sub{BZ}\frac{\D\vec{k}}{(2\pi)^3} \sum_n 
	\frac{\partial f_{\vec{k}n}}{\partial\varepsilon_{\vec{k}n}} |\vec{v}_{\vec{k}n}|^2
},
\label{eqn:tauD}
\end{equation}
which is just an average of the per-state momentum relaxation time near the Fermi level, weighted
by the band velocity squared because that determines the contribution to the conductivity above.

\begin{table}
\caption{Comparison of calculated resistivity of transition metal nitrides
against measured values where available. $\tau_\mathrm{D}$ is the Drude lifetime in femtoseconds and v$_\mathrm{F}$ is Fermi velocity.  
\label{tab:resistivity}}
\begin{tabular}{|c|c|c|c|c|c|}
\hline
\multirow{2}{*}{TMN} & \multirow{2}{*}{Phase} & \multirow{2}{*}{$\tau_D$[fs]}
& \multirow{2}{*}{$v_F$[10$^{6}$m/s] } &\multicolumn{2}{c|}{Resistivity, $\rho$ [n$\Omega$m]} \\
\cline{5-6}
& & &   & Calc & Expt (S) \\
\hline
TiN & RS & 6.9 & 0.72 & 123 & 129\cite{e-ph-IV}  \\
\hline
ZrN & RS & 5.8 & 1.01 & 114 & 120\cite{e-ph-IV} \\
\hline
HfN & RS & 5.4 & 1.08 & 112 & 142\cite{e-ph-IV} \\
\hline\hline
VN  & Hex & 99. & 0.28 & 191 &  \\
\hline
NbN & Hex & 68. & 0.36 & 381 &  \\
\hline
TaN & Hex & 37. & 0.73 & 139 &  \\
\hline\hline
\multirow{2}{*}{CrN} 
& Hex & 17.6  & 0.33 & 99  &  \\
& AFM & 9.5 & 0.47 & 351 &  \\
\hline
\multirow{2}{*}{MoN}
 & Hex & 4.2 & 0.63 & 220 &  \\
 & NbO & 6.9 & 0.53 & 265 &  \\
\hline
\multirow{2}{*}{WN}
 & Hex & 3.0 & 0.72 & 249 &  \\
 & NbO & 8.9 & 0.67 & 190 &  \\
\hline
\end{tabular}
\end{table}

Table~\ref{tab:resistivity} reports the calculated resistivity values, average Fermi velocities and
Drude relaxation time for all mechanically stable structures of the nine TMN materials considered here.
For the group IV nitrides, where RS structure is theoretically stable at zero temperature
and also experimentally stable at room temperature, the calculated resistivities are in excellent
agreement with experimental measurements on single-crystalline / epitaxially-grown materials \cite{e-ph-IV}.
However, for all other nitrides, experimental values in the literature span several orders of magnitude,
ranging from $10^3 - 10^6$~n$\Omega$m, while the theoretical predictions remain on the same order of magnitude
as the group IV nitrides. This is because these materials tend to adopt primarily non-stoichiometric
configurations with very large concentrations of defects that substantially increase carrier scattering.

In comparison, note that silver has a resistivity of 16~n$\Omega$m, a Drude relaxation
time of 35~fs and a Fermi velocity $\sim 1.4\times 10^6$~m/s \cite{PhononAssisted}.
Therefore, the nitrides have resistivities that are typically one order of magnitude higher,
but they differ drastically in the combination of factors that lead to that result.
The cubic-based structures (RS, AFM and NbO) all have shorter relaxation times
and Fermi velocities comparable to silver, while the hexagonal structures have
comparable or longer relaxation times than silver, but smaller Fermi velocities.
This, in turn, is caused by a lower density-of-states at the Fermi level for
the hexagonal nitrides, which lowers the rate of electron-phonon scattering.
In particular, we predict hexagonal VN to exhibit a momentum relaxation time
that is almost three times larger than silver!

\subsection{Optical response and carrier generation}

\begin{figure}
\includegraphics[width=\columnwidth]{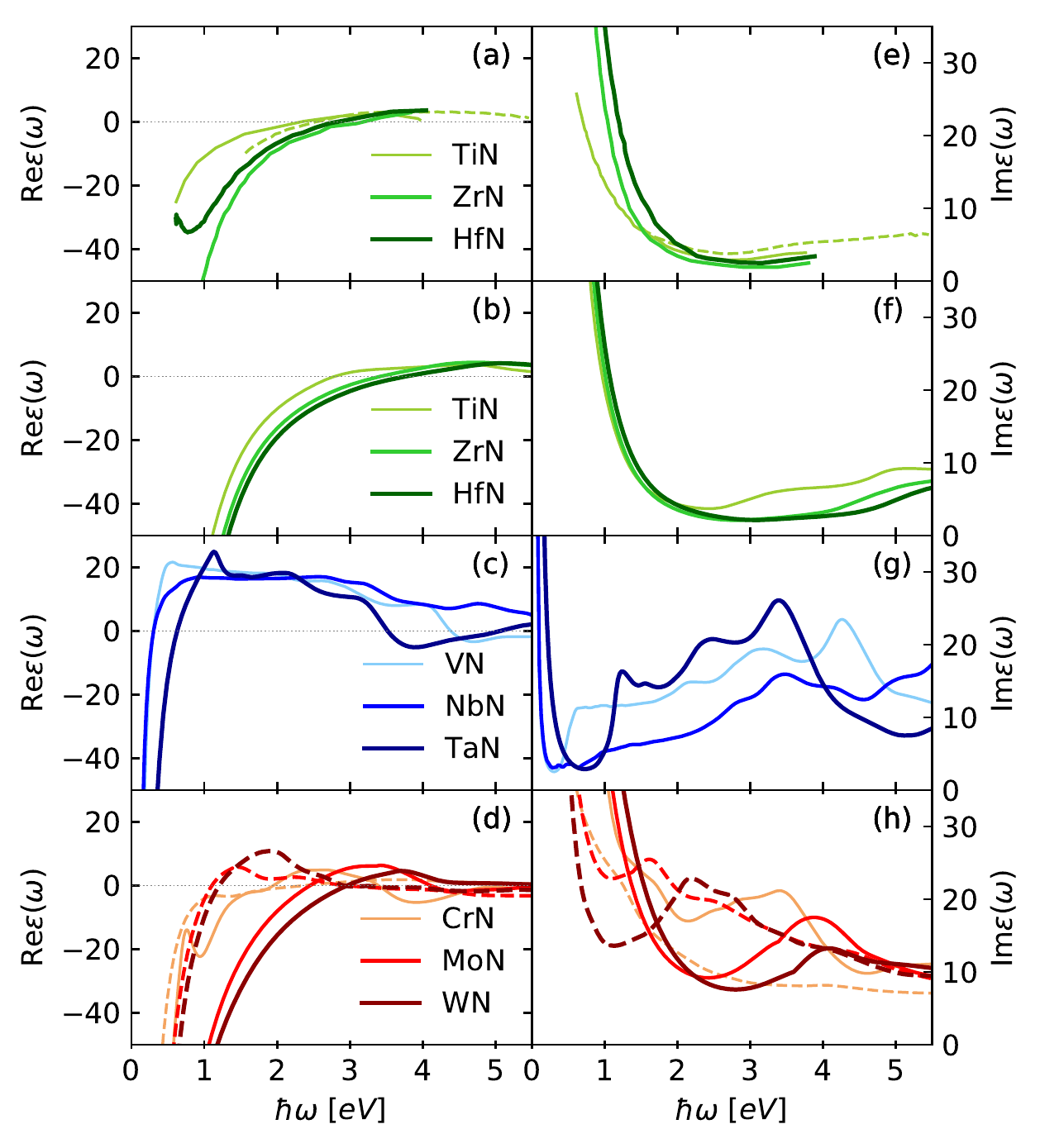}
\caption{(a,e) Measured dielectric functions for
the Ti-group nitrides \cite{NaikEpsData,PatTiNEpsData} compared to
\emph{ab initio} predictions of the dielectric functions
including direct interband, phonon-assisted and Drude intraband contributions
from first principles for nitrides of (b,f) Ti, (c,g) V and (d,h) Cr
group metals respectively, with real parts in the left panels
and imaginary parts in the right panels.
In (a,e), solid lines are experimental data from \cite{NaikEpsData},
and the dashed line for TiN is from \cite{PatTiNEpsData}.
In (d,h), solid lines represent the hexagonal structure for each material,
while the dashed lines represent the tetragonal AFM structure for CrN
and the NbO structure for MoN and WN.
\label{fig:epsilon}}
\end{figure}

Next, we calculate the complex dielectric function for each
of the mechanically-stable TMN structures as
\begin{equation}
\bar{\epsilon}(\omega) = 1 + \frac{4\pi i\bar{\sigma}}{\omega(1-i\omega\tau_D)} + \bar{\epsilon}_d(\omega) + \bar{\epsilon}_p(\omega),
\end{equation}
consisting of the Drude intraband term with $\sigma$ and $\tau_D$ calculated in
the previous section, and contributions from direct and phonon-assisted transitions.
From Fermi's Golden rule, the imaginary part of the direct (interband) contribution is
\begin{multline}
\hat{\lambda}\cdot\Im\bar{\epsilon}_d(\omega)\cdot\hat{\lambda}
= \frac{4\pi^2 e^2}{\omega^2} \int\sub{BZ} \frac{d\vec{k}}{(2\pi)^3} \sum_{n'n}
	(f_{\vec{k}n} - f_{\vec{k}n'})
\\\times
	\delta(\varepsilon_{\vec{k}n'} - \varepsilon_{\vec{k}n} - \hbar\omega)
	\left| \hat{\lambda} \cdot \vec{v}^{\vec{k}}_{n'n} \right|^2,
\label{eqn:ImEpsDirect}
\end{multline}
while that of the phonon-assisted (intraband) contribution is
\begin{multline}
\hat{\lambda}\cdot\Im\bar{\epsilon}_p(\omega)\cdot\hat{\lambda}
= \frac{4\pi^2 e^2}{\omega^2}
	\int\sub{BZ} \frac{d\vec{k}'d\vec{k}}{(2\pi)^6} \sum_{n'n\alpha\pm}
	(f_{\vec{k}n} - f_{\vec{k}'n'})
\\\times
	\left( n_{\vec{k}'-\vec{k},\alpha} + \frac{1}{2} \mp \frac{1}{2}\right)
	\delta(\varepsilon_{\vec{k}'n'} - \varepsilon_{\vec{k}n} - \hbar\omega \mp \hbar\omega_{\vec{k}'-\vec{k},\alpha}) \\
\times \left| \hat{\lambda} \cdot \sum_{n_1} \left(
	\frac{ g^{\vec{k}'-\vec{k},\alpha}_{\vec{k}'n',\vec{k}n_1} \vec{v}^{\vec{k}}_{n_1n} }{ \varepsilon_{\vec{k}n_1} - \varepsilon_{\vec{k}n} - \hbar\omega + i\eta}
\right. \right. \\
+ \left. \left.
	\frac{ \vec{v}^{\vec{k}'}_{n'n_1} g^{\vec{k}'-\vec{k},\alpha}_{\vec{k}'n_1,\vec{k}n} }{ \varepsilon_{\vec{k}'n_1} - \varepsilon_{\vec{k}n} \mp \hbar \omega_{\vec{k}'-\vec{k},\alpha} + i\eta }
\right) \right|^2,
\label{eqn:ImEpsPhonon}
\end{multline}
where all the energies, occupation factors and matrix elements are as defined above
and evaluated form a Wannier representation, and where $\hat{\lambda}$ is a test unit
vector along the field direction to conveniently express the tensorial result.
The corresponding direct parts are evaluated by the Kramers-Kronig relation.
The energy-conserving $\delta$-functions are broadened to Lorentzians with width
equal to the sum of the initial and final carrier linewidths ($\Im\Sigma$), which are
also evaluated \emph{ab initio} as indicated in the carrier transport section below,
and $\eta$ is used for eliminating singular contributions in the phonon-assisted result.
By histogramming integrands in the above expressions with respect to $\epsilon_{\vec{k}n}$,
we also evaluate the energy distributions of carriers generated upon absorption.
See Ref.~\citenum{PhononAssisted} for further details and demonstration of the
quantitative accuracy of this method for conventional plasmonic metals.

\begin{figure}
\includegraphics[width=\columnwidth]{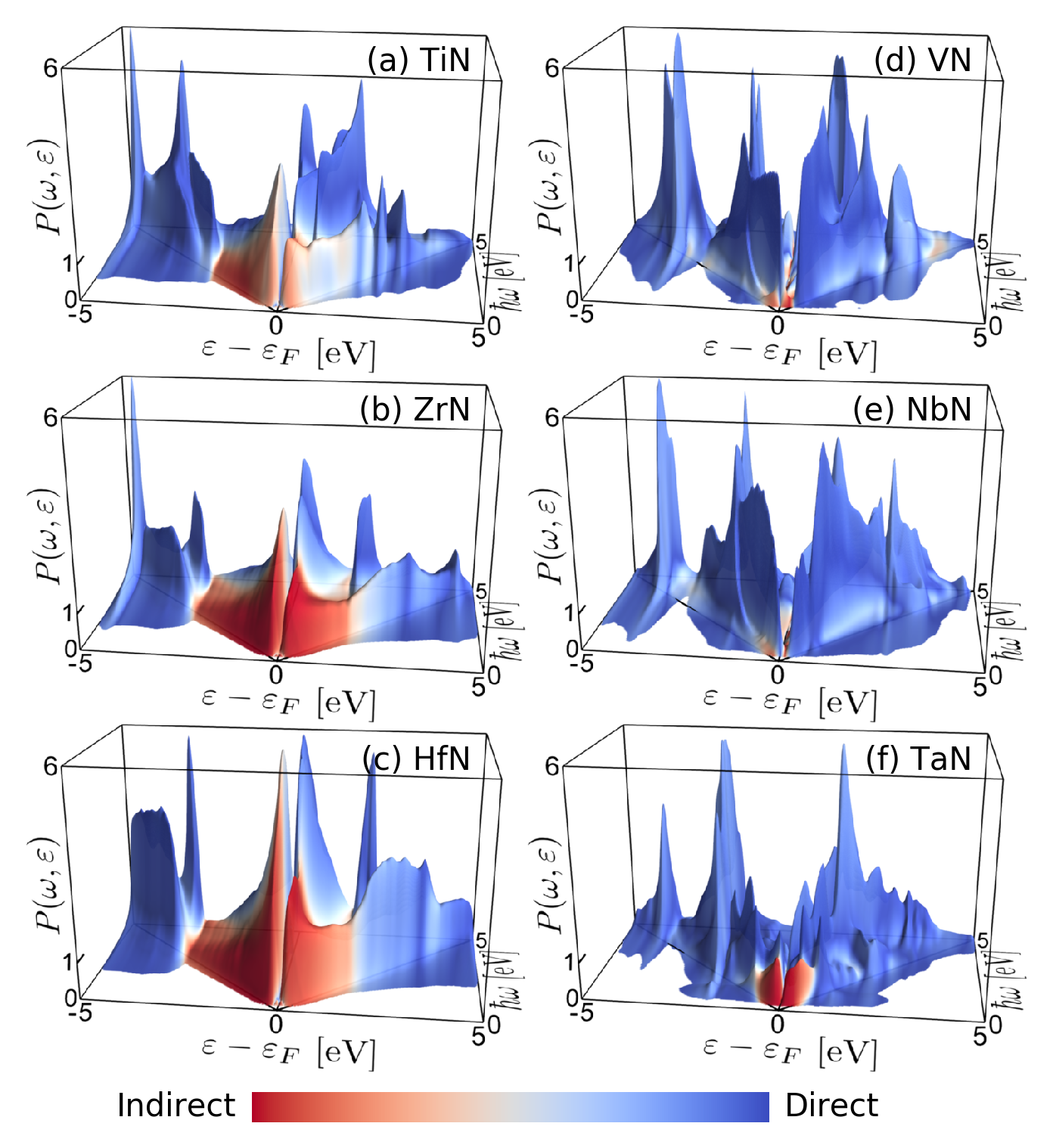}
\caption{Energy distribution of hot carriers generated as
a function of plasmon/photon frequency ($\hbar\omega$)
in (a-c) Ti group nitrides and (d-f) V-group nitrides.
The probability $P(\omega,\varepsilon)$ is normalized such that 
a uniform carrier distribution is 1.
The color scale indicates relative contribution of intraband processes:
intraband processes dominate only at low energies, particularly for the
Ti-group nitrides in the rock-salt structure, while interband processes dominate
hot carrier generation for most of the materials and shown energy range.
\label{fig:carrierDistribA}}
\end{figure}

\begin{figure}
\includegraphics[width=\columnwidth]{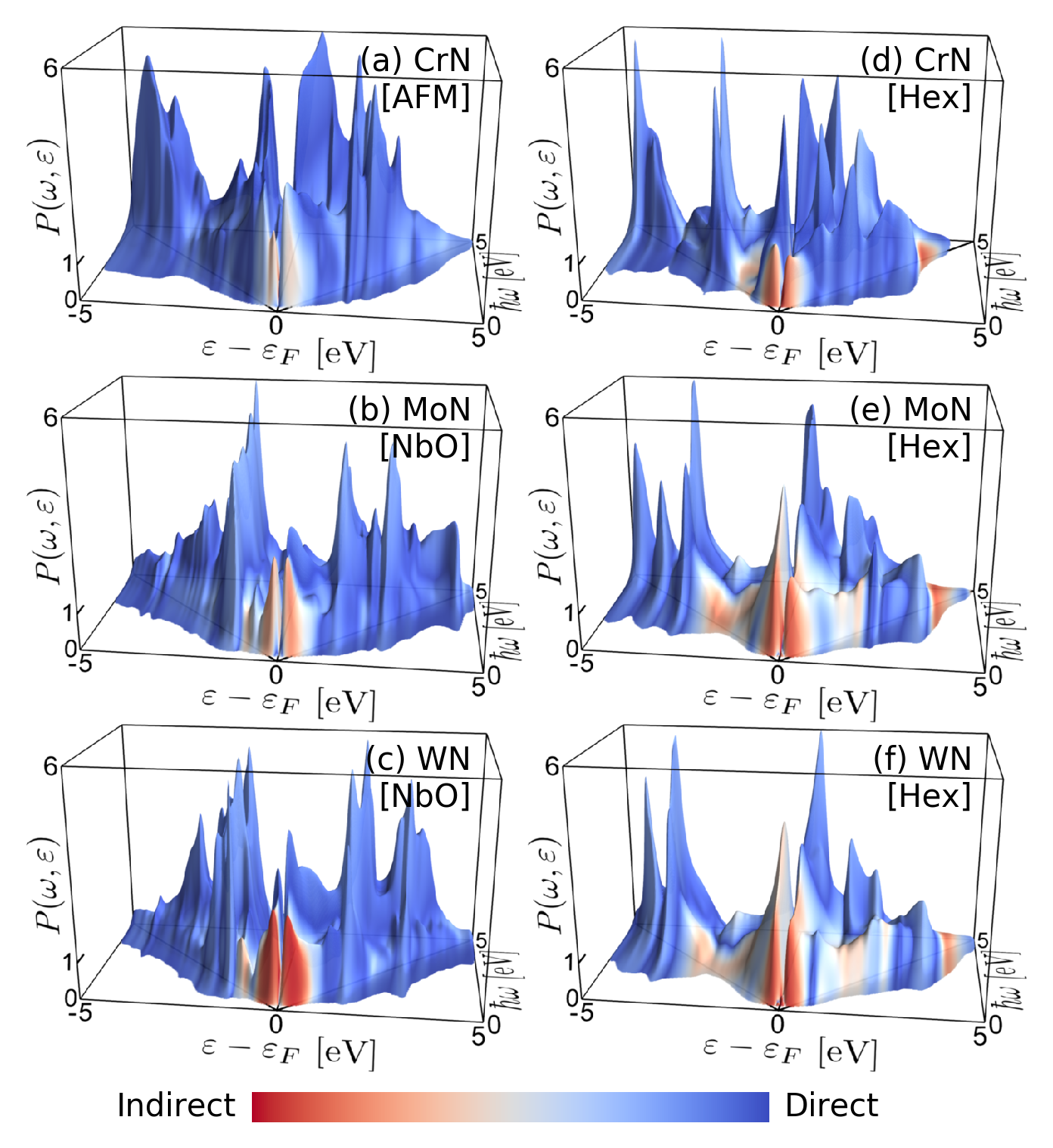}
\caption{Same as Fig.~\ref{fig:carrierDistribA} for the Cr-group nitrides (a) in the tetragonal AFM structure, (b-c) in the NbO structures, and (a-c) in the hexagonal structures.
\label{fig:carrierDistribB}}
\end{figure}

Figure \ref{fig:epsilon} shows the predicted dielectric functions for the stable TMN phases
The RS group IV nitrides and the Hex group VI nitrides all exhibit a wide energy window
of negative $\Re\epsilon$ corresponding to plasmonic behavior,
ranging from 2~eV for CrN to almost 4~eV for HfN.
These windows are comparable to about 2.6~eV for gold and 3.4~eV
for silver calculated using the same method \cite{TAparameters};
the experimental values for the noble metals are about 0.2 - 0.3~eV larger.
The corresponding $\Im\epsilon$ where $\Re\epsilon$ crosses zero is also
quite low, indicating excellent plasmonic behavior over this energy range
extending across the visible region and into the near UV.
In contrast, group V nitrides and the NbO/AFM-structure group VI
nitrides exhibit a much narrower plasmonic window, less than 1~eV.
However, within that narrow window, the group V nitrides exhibit
substantially lower $\Im\epsilon$ due to the much longer Drude
relaxation time discussed in the previous section.

The calculated dielectric functions are in good agreement 
with experimental data for the group IV nitrides in RS structure
\cite{NaikEpsData,PatTiNEpsData} shown in Fig.~\ref{fig:epsilon}(a,e).
In particular, the energy at which $\Re\epsilon$ crosses zero,
demarcating the plasmonic window for these metals, agrees with
experiment within 0.2~eV for TiN and ZrN, and within 1~eV for HfN.
This agreement is well within the sample to sample variation of measured
optical properties of TMNs, which are extremely sensitive to stoichiometry,
impurities and growth methods \cite{NaikEpsData,PatTiNEpsData}.
These issues are particularly severe for group V and VI TMNs
\cite{PlasmonicNitridesReview,WNstabilityGall} and comparison
to experiment for those materials require explicit treatment
of defective systems, a subject of ongoing work.
Finally, note that the calculated dielectric functions of the group IV nitrides are in
good agreement with previous calculations that treated the intraband contributions empirically
\cite{PlasmonicResponseTheoryTMN}; see Ref.~\citenum{PlasmonicResponseTheoryTMN} for a
detailed comparison of the predicted plasmonic properties of these nitrides with noble metals.

Here, we instead focus on the plasmonic hot carrier properties of these
nitrides, where plasmonic loss is not an issue, but rather the design goal.
Fig.~\ref{fig:carrierDistribA} and \ref{fig:carrierDistribB} show the
predicted energy distributions of generated hot carriers for each material
as a function of plasmon/photon energy, and also indicates the relative
contributions of phonon-assisted and direct transitions.
The group IV nitrides with wide plasmonic windows generate carriers
predominantly by phonon-assisted transitions at lower photon energies,
which then switches to predominantly direct transitions above an
interband threshold, analogously to the noble metals \cite{PhononAssisted}.
All remaining nitrides support direct transitions at almost all
energies and the dominance of phonon-assisted transitions is narrow,
roughly parallel to the plasmonic windows.

Additionally, in contrast to the noble metals, the carrier distributions
in all cases, direct or indirect, exhibit complex features due to the
more complicated band structures and do not clearly have regions of
uniform carrier energy distributions as the noble metals do \cite{PhononAssisted}.
While the carrier distributions are `spiky', they appear to cover
the entire energy range extending from photon energy below the
Fermi level to photon energy above it (the small contributions beyond
the photon energy are due to carrier broadening, and are allowed
for these short-lived carriers by the uncertainty principle).
However, they are not perfectly symmetrical between electrons and holes,
just as in the noble metals, and it is important to know if the 
energy distributions are electron or hole-dominant in order
to decide suitability for $n$-type collection / reductive photochemistry
or $p$-type collection / oxidative photochemistry respectively.

\begin{figure}
\includegraphics[width=\columnwidth]{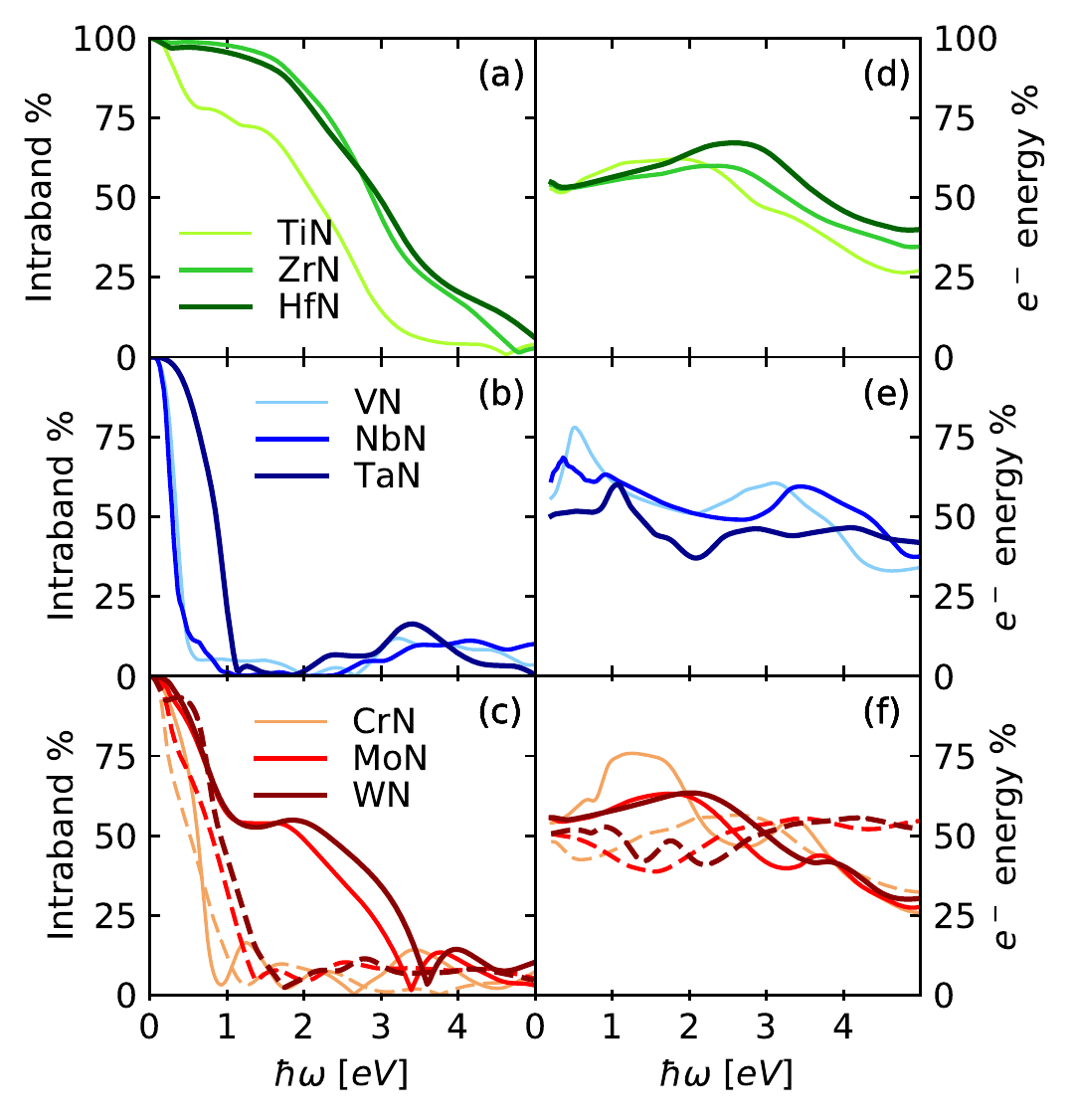}
\caption{(a,b,c) Fraction of hot carrier generation due to
indirect intraband processes for each of the transition metal nitrides.
(d,e,f) Corresponding fraction of average energy deposited in hot electrons,
where 50\% corresponds to equal distribution of energy between electrons and holes.
In (c,f), solid lines represent the hexagonal structure for each material,
while the dashed line represents the tetragonal AFM structure for CrN
and the NbO structure for MoN and WN.
Intraband processes dominate with equal electron and hole energies at low frequencies,
while interband processes at higher frequencies slightly bias carrier energies towards holes.
\label{fig:carrierDistribStats}}
\end{figure}

Figure~\ref{fig:carrierDistribStats} summarizes the relative contribution from
intraband transitions and energy dominance of electrons versus holes for the TMNs.
The cross-over from intraband to direct transitions on the left panels correlates very
well with the end of the plasmonic window ($\Re\epsilon<0$ in Fig.~\ref{fig:epsilon}).
Correspondingly direct (interband) carrier generation becomes important for the RS
group IV nitrides, Hex MoN and Hex WN above 3~eV, and above 1~eV for all other nitrides.
The indirect (intraband) processes distribute energy between hot electrons and holes
roughly evenly, while direct transitions tend to bias the carrier distribution towards
holes, just as for the noble metals \cite{DirectTransitions}, but to a lesser extent.
The underlying reason for this remains that these direct transitions are predominantly
from metal $d$-bands below the Fermi level to just above the Fermi level.
Also, interestingly the energy bias swings slightly towards the
electrons for photon energies just below the interband threshold
(indicated by the humps in Fig.~\ref{fig:carrierDistribStats}(d-f)),
which arises because of the (small) asymmetry in the conduction band
density of states from below the Fermi level to above it.
Altogether, all these nitrides are capable of generating both
hot electrons and holes with distributions extending mostly all the
way to the photon energy, almost irrespective of the photon energy.

\subsection{Carrier transport}

Finally, we calculate the lifetimes and mean-free paths of hot carriers
generated in these materials which are the determining factors for
the efficiency of plasmonic hot carrier applications.
The electron-phonon contribution to the inverse lifetime is
\begin{multline}
\left(\tau\sub{e-ph}^{-1}\right)_{\vec{k}{n}} =
\frac{2\pi}{\hbar} \int\sub{BZ} \frac{\Omega d\vec{k}'}{(2\pi)^3} \sum_{n'\alpha\pm}
	\delta(\varepsilon_{\vec{k}'n'} - \varepsilon_{\vec{k}n} \mp \hbar\omega_{\vec{k}'-\vec{k},\alpha})
\\
\times
	\left( n_{\vec{k}'-\vec{k},\alpha} + \frac{1}{2} \mp \left(\frac{1}{2} - f_{\vec{k}'n'}\right)\right)
	\left| g^{\vec{k}'-\vec{k},\alpha}_{\vec{k}'n',\vec{k}n} \right|^2,
\label{eqn:tauInv_ePh}
\end{multline}
identical to the momentum relaxation rate given by (\ref{eqn:tauInv_ePhP}),
except without the factor accounting for the angle between initial and final velocities.
The electron-electron contribution is calculated using
\begin{multline}
\left(\tau\sub{e-e}^{-1}\right)_{\vec{k}{n}} =
\frac{2\pi}{\hbar} \int\sub{BZ} \frac{d\vec{k}'}{(2\pi)^3} \sum_{n'}
\sum_{\vec{G}\vec{G}'}
	\tilde{\rho}_{\vec{k}'n',\vec{k}n}(\vec{G})
	\tilde{\rho}_{\vec{k}'n',\vec{k}n}^\ast(\vec{G}')\\
\times \frac{1}{\pi}\Im\left[ \frac{4\pi e^2}{|\vec{k}'-\vec{k}+\vec{G}|^2}
	\epsilon^{-1}_{\vec{G}\vec{G}'}(\vec{k}'-\vec{k},\varepsilon_{\vec{k}n}-\varepsilon_{\vec{k}'n'}) \right],
\label{eqn:tauInv_ee}
\end{multline}
where $\tilde{\rho}_{\vec{q}'n',\vec{q}n}$ are density matrices
in the plane-wave basis with reciprocal lattice vectors $\vec{G}$,
and $\epsilon^{-1}_{\vec{G}\vec{G}'}$ is the inverse dielectric matrix
calculated in the Random Phase Approximation.
See Ref.~\citenum{eeLinewidth} for a detailed introduction to this method
and Ref.~\citenum{PhononAssisted} for our specific implementation details.
The net carrier lifetime is $\tau^{-1}_{\vec{k}n} = (\tau\sub{e-ph}^{-1})_{\vec{k}n}
+ (\tau\sub{e-e}^{-1})_{\vec{k}n}$, and from it we calculate the carrier linewidths
$\Im\Sigma_{\vec{k}n} = \hbar/(2\tau_{\vec{k}n})$ for the broadening in
(\ref{eqn:ImEpsDirect}) and (\ref{eqn:ImEpsPhonon}) above, as well as 
the mean free path $\lambda_{\vec{k}n} = |\vec{v}_{\vec{k}n}| \tau_{\vec{k}n}$.

\begin{figure}
\includegraphics[width=\columnwidth]{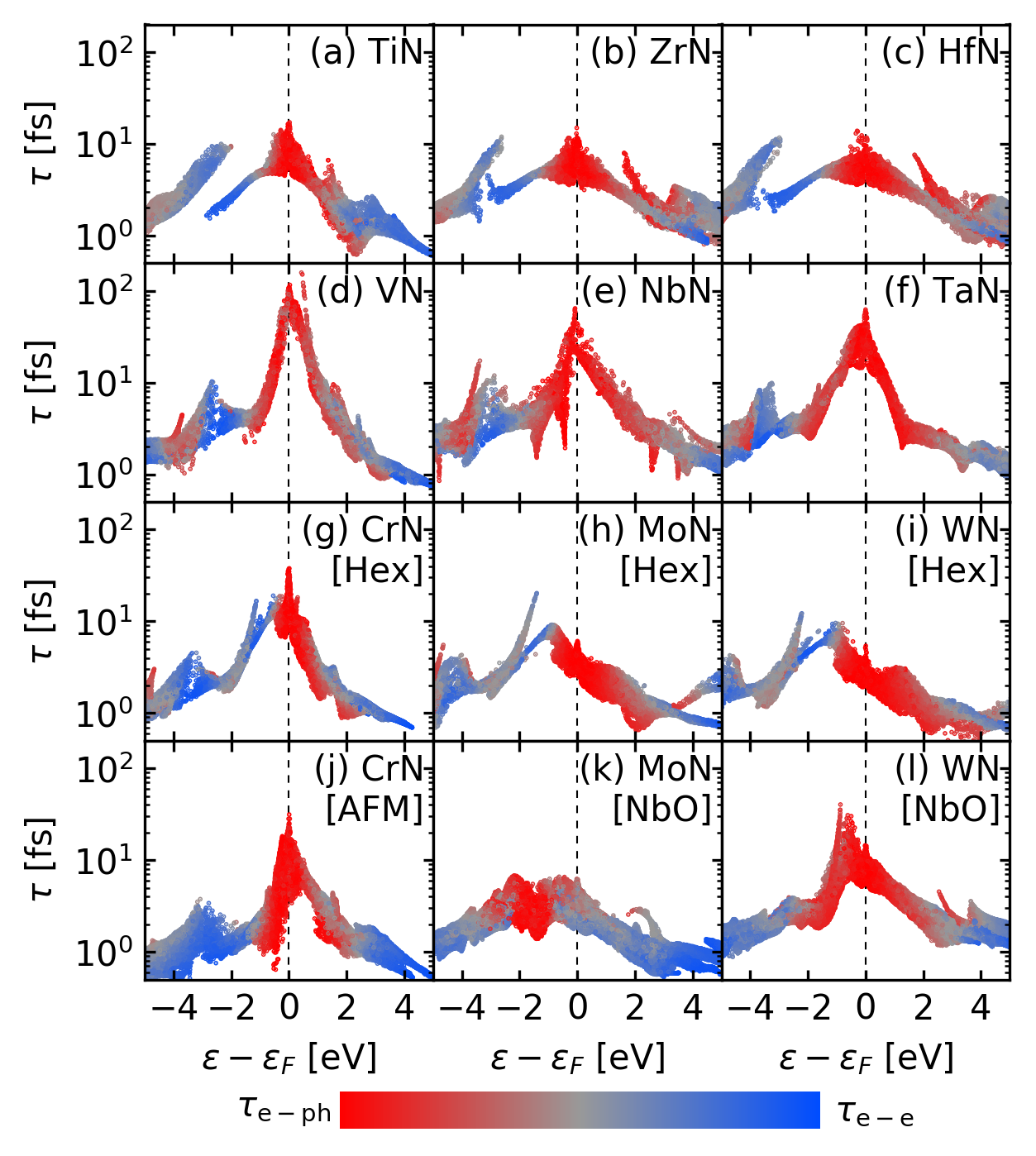}
\caption{Hot carrier lifetimes as a function of energy, accounting for electron-electron and electron-phonon contributions for each of the transition metal nitrides.
Horizontally, (a-c) are group Ti TMNs in rock salt, (d-f) are group V TMNs in hexagonal, (g-i) are group Cr TMNs in hexagonal, (j) is CrN in tetragonal AFM,
and (k-l) are MoN and WN in NbO structures. The color scale shows relative contributions of electron-electron (blue) to electron-phonon (red) scattering rates to the lifetime.
\label{fig:tau}}
\end{figure}

\begin{figure}
\includegraphics[width=\columnwidth]{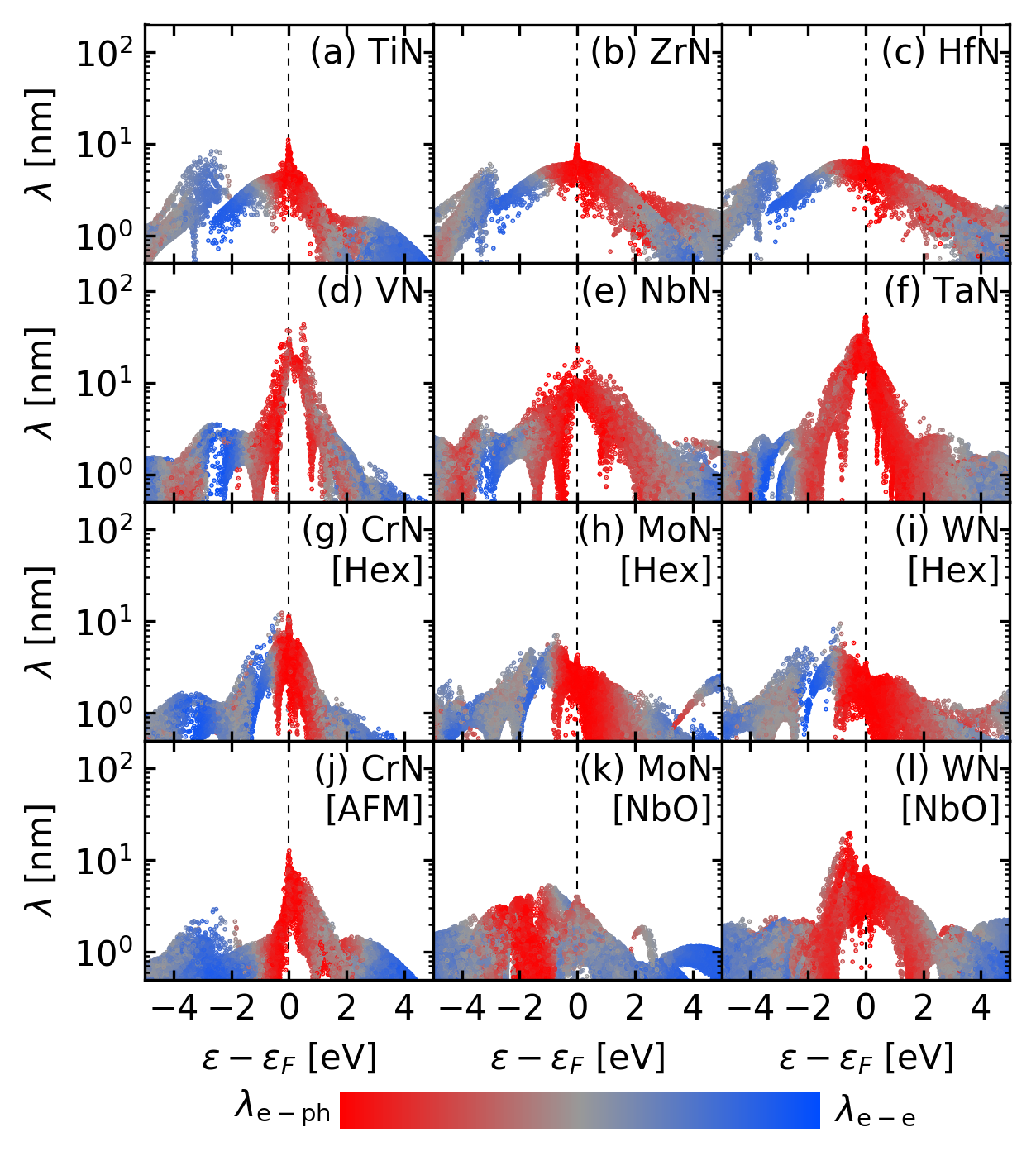}
\caption{Same as Fig.~\ref{fig:tau}, but for the mean free path instead of lifetime.
\label{fig:lambda}}
\end{figure}

Figures~\ref{fig:tau} and \ref{fig:lambda} respectively show the calculated lifetimes and
mean free paths for all TMNs as a function of carrier energy away from Fermi level, indicating
relative contributions of electron-electron and electron-phonon processes using a color scale.
In each panel of both figures, the maximum lifetime (or mean free path) typically occurs near
the Fermi level, where it is dominated by electron-phonon scattering and where the phase space
for electron-electron scattering is negligible.
The lifetime rapidly diminishes with increasing energy away from the Fermi level due
to the increased phase space for electron-electron scattering, clearly indicated
by the change from red to blue in the figures.
This behavior is qualitatively similar to that of the noble metals \cite{PhononAssisted},
except that the shapes of the fall-off are more complicated here on account of more
complex density of states profiles compared to the free-electron-like ones in the noble metals.
Sharp discontinuities seen in Fig.~\ref{fig:tau}(a-c,h,i) correspond to new bands opening up
at that energy, and states near that band edge have a long life-time due to reduced scattering phase space.
However, the corresponding panels in Fig.~\ref{fig:lambda} do not exhibit these discontinuities,
or strongly suppress them, because carrier velocities near these band edges are small.

Figures~\ref{fig:tau}(d-f) show that group V TMNs have
the highest life times for carriers near the Fermi level,
almost an order of magnitude greater than most of the other nitrides.
In fact, just as was the case for the Drude relaxation time
(Table~\ref{tab:resistivity}), NbN and TaN have comparable lifetimes to silver,
the highest among noble metals, and VN exceeds it by a factor of two.
Among the remaining nitrides, WN in the NbO phase and the two
CrN structures exhibit the highest lifetimes, approaching 20-30 fs.
The corresponding advantage for group V nitrides over the rest
in terms of mean free paths (Fig.~\ref{fig:lambda}(d-f)) is less striking,
because of an overall lower band velocity near the Fermi level.
As discussed above, this also resulted in similar resistivities compared
to other nitrides despite substantially longer relaxation times.
In terms of mean free path, TaN leads with maximum values exceeding 40~nm,
with NbN and VN close behind at around 30~nm.
However, the lifetimes drop below 10~fs and mean free paths below 10~nm
for carrier energies exceeding 1~eV across the board, similar to what
we found for graphene and its heterostructures \cite{GraphiteHotCarriers}.
The nitrides fall behind silver in this regard, where the carrier lifetimes
and mean free paths decline more gradually with energy from the Fermi level \cite{PhononAssisted}.

\section{Conclusions}

Using \emph{ab initio} calculations of optical response, hot carrier generation
and carrier transport, we show that plasmonic transition metal nitrides are
promising candidates for efficient hot carrier extraction.
Group IV and VI nitrides exhibit wider frequency windows for plasmonic response,
while group V nitrides exhibit the best carrier transport properties.
Our completely parameter-free predictions for the resistivity of these materials
are in excellent agreement with experimental results on stoichiometric nitrides,
currently available for group IV nitrides.

Hot carrier generation is dominated by direct transitions for optical frequencies
in most of the nitrides, and intraband transitions become important primarily
at infrared frequencies for the group IV nitrides.
Generated carrier energy distributions tend to be more evenly distributed between
electrons and holes compared to noble metals, and are only slightly biased
towards holes when interband transitions dominate.

Finally, in terms of carrier transport, hexagonal VN exhibits the highest
carrier lifetime, far exceeding silver, but on account of the lower Fermi
velocity in VN, TaN exhibits the highest carrier mean free path, comparable to silver.
High-energy ($>1~eV$) carrier transport in these materials is somewhat inferior
to the noble metals, but an important advantage of these refractory plasmonic
materials remains that they can be stabilized at much smaller dimensions,
potentially circumventing limitations due to shorter carrier transport distances.

A key bottleneck in current \emph{ab initio} studies of hot carrier dynamics
is the limitation to structures mechanically stable at zero temperature, inherent
because of the perturbative nature of the calculations relative to the perfect crystal.
This precluded calculations of rock-salt structures of materials such as VN,
which are mechanically stable at room temperature but unstable at zero temperature.
More generally, phonon-related properties of TMNs could be sensitive
to temperature even when the room-temperature structures are stable at zero temperature.
The capability to account for finite temperature effective force matrices and electron-phonon
matrix elements from \emph{ab initio} molecular dynamics simulations is therefore
an important future direction in theoretical calculations of hot carrier dynamics
in general, and to plasmonic nitrides in particular.

\section*{Supplemental Information}

The supplemental information includes plots of electronic band structures,
density of states, phonon band structures, and phonon density of states
for all nitride structures considered above.

\section*{Acknowledgments}
RS acknowledges startup funding from the Department of Materials Science
and Engineering at Rensselaer Polytechnic Institute.
Calculations were performed on the BlueGene/Q supercomputer in the
Center for Computational Innovations (CCI) at Rensselaer Polytechnic Institute.

\bibliographystyle{apsrev4-1}
\makeatletter{} 
\end{document}